\documentclass[onecolumn,preprintnumbers,amsmath,amssymb]{revtex4}
\usepackage{graphicx}
\usepackage{dcolumn}
\usepackage{bm}
\begin{document}

\title{Introduction to Critical Phenomena through the Fiber Bundle Model of Fracture}
\author{Srutarshi Pradhan\email{srutarshi.pradhan@ntnu.no}} 
\affiliation{PoreLab, Department of Physics, Norwegian University of
Science and Technology, NO--7491 Trondheim, Norway.}
\author{Bikas K. Chakrabarti\email{bikask.chakrabarti@saha.ac.in}}
\affiliation{Saha Institute of Nuclear Physics, Kolkata 700064, India} 
\affiliation{S. N. 
Bose National Centre for Basic Sciences, Kolkata 700106, India.}

\begin{abstract}
We discuss the failure dynamics of the Fiber
Bundle Model, especially in the equal-load-sharing scheme.
We also highlight the ``Critical" aspects of their dynamics in
comparison with those in standard thermodynamic systems undergoing phase
transitions.
\end{abstract}
\maketitle
\section{Introduction}
After receiving the invitation from the Editors to
contribute to the special issue on ``Complexity" of the
European Journal of Physics, we thought that it would
be a good idea to introduce the readers, students in
particular, to the study of the dynamics of the ``Fiber
Bundle Model" (FBM) and to their ``critical behavior".
We understand that 
the main focus of this special issue is to present the topic of 
``Complexity" in a simple way, demonstrating the subtle features of
the problems.
``Critical Phenomena" nicely demonstrate one important aspect of the 
complexity of the nature
and this is  wide enough a field with huge developements in all the fronts:
 theoretical, numerical and experimental. Thousands
of papers, reviews and articles have been published  so far on this topic 
and few
hundreds are still being published each year. But introducing 
critical phenomena to the college, university students or to the
beginners is not always easy. Surely,  asking them to go through the vast
literature  does not make sense.

Here comes the role of ``models".
We plan to introduce them through a ``simple" model which is intuitively 
appealing and has clear-cut dynamical rules.
They allow us to perform some analytic calculations and get solutions. Also
it is possible to check the results numerically through a few quick
``runs".

The Fiber Bundle Model (FBM), introduced by Peirce in $1926$ \cite{p26}, 
is such a simple model. Although quite old
and even though it was designed as a model for
fracture or failure
of a set of parallel elements (fibers), each having a breaking threshold 
different from others, with the collective sharing of the failed fiber's load, 
 the failure dynamics in the model clearly shows
all the attributes of the critical phenaomena and associated phase
transition. Indeed, the FBM 
is the precise equivalent of the Ising model of
magnetism, introduced by Ising \cite{i25} in 1925. The models
are both identically potential, deep and versatile in
the respective fields.
A recently published book \cite{hhp15} tried to gather and explain
works on FBM from a ``statistical physics" point of view. 
In this article we want to concentrate only on FBM's dynamical
aspects related to ``critical phenomena". We hope, it will help easy 
introduction to the models (FBM) and critical phenomena before the students 
undertake research on these topics. 
Yes, ``students" are the target groups for this article.

We arrange this article as follows: After the short introduction
(section I) we briefly introduce the notion of critical phenomena 
in section II. Then we
discuss the brief history of FBM and its evolution as a
fracture model in section III. Section IV deals with the simplest version
of FBM, the equal-load-sharing FBM. In several sub-sections of
section IV we demonstrate the critical behavior in FBM with
construction of the evolution dynamics and their solutions. We also
compare the analytic results with numerical simulations in this
 section. We dedicate section V for discussions on some related works which 
would help
to understand the critical behavior of FBM. Finally, some discussions
and conclusions will be made in section VI.

\section{Critical phenomena}
Let us consider the case of a ferromagnet. Even  in absence of any external 
field, at temperatures ($T$) below the Courie temperature ($T_c$), one gets a 
 finite 
average magnetisation. This spontaneous 
magnetisation disappeares as one increases the temperature of the magnet 
beyond the Curie value. This phase transition (from the ferromagnetic 
 phase with spontaneous magnetisation to paramagnetic phase with vanishing 
magnetisation) was seen to have some ``critical" aspects in the sense that the
temperature variation of the magnetisation ($m$) or the susceptibility ($\chi$) 
near the 
Curie point can be expressed as power laws with respect to the temperature 
interval from the Curie (critical) point: $m(T) \sim |T-T_c|^{\beta}$, 
$\chi(T) \sim |T-T_c|^{-\gamma}$. Additionally, the values of these 
 powers
(exponents) were found to be irrational numbers in general (indicating 
singular behavior of the free energy of the magnets near the ferro-para 
transition point). When interpreted in terms of the elemental spin-magnetic 
moments, which interact through the exchange interactions, one finds that the 
correlation of the spin-state fluctuations at any arbitrary crystal point and 
that of another spin 
at a 
distance $r$ decayes as $exp[-r/\xi(T)]$, where the correlation length 
 $\xi(T) \sim |T-T_c|^{-\nu}$ diverges at 
$T=T_c$ with correlation length exponent $\nu$. This correlation length sets 
the scale which essentially determines the critical aspects of the 
thermodynamic behavior near the crirical (Curie) point of the magnet.
One also finds values of these exponents ($\beta$, $\gamma$, $\nu$ etc.) are 
universal in the sense that they depend only on some subtle featurs of the 
systems like spatial dimensionality of the system and the number of 
components (dimesionality) of the order parameter (magnetisation) vector.  
They do not depend on the details like strength of (exchange) interaction, 
lattice structures, etc.      

In the standard models of cooperatively interacting systems
in classical statistical physics, like in the Ising model,
simple two-state Ising spins (representing the constituent magnetic moments) 
on the lattice sites interact
with themselves through neighboring (exchange) interactions. In absence
of any thermal noise, or even at finite but low temperatures,
the effect of the (spin-spin) interactions win over thermal
noise, and induce spontaneous order without any magnetic
field. This order, say ferromagnetic when the spin-spin
interactions favor similar orientations of the spins, gets
destroyed when the thermal noise, corresponding to a
temperature beyond the phase transition or Curie point, wins
over and paramagnetic phase sets in. After intensive studies
for about three decades, starting middle of the last
century, it was established that while the thermodynamic
behavior away from the phase transition point of such
systems have the usual scale dependent variations (with
the finite scale determined by the competitions between the
interaction energies and  temperature), the behavior become
scale free as the phase transition point is approached (see
e.g., \cite{f17})  and the behaviors are expressed by power laws
(with the powers given by some effective fractal
dimensionality \cite{m82} of ``volume" determined by the ``correlation
length" scale which diverges at the phase transition point).
This scale invariance, often with singularities in the
growth of the correlation length scale as one approaches
the transition point, had been exploited by the renormalization
group theory (see e.g., \cite{f17,m82,wk74}).

\section{The Fiber Bundle Model}
\begin{figure}[t]
\begin{center}
\includegraphics[width=6cm]{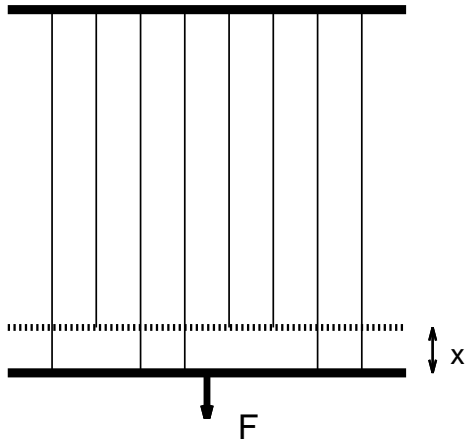}
\vspace{0.6cm}
\caption {\label{fig:FBM-model}
The Fiber Bundle Model: fibers in parrallel are placed between two solid bars 
and the load is applied at the lower bar.}
 
\end{center}
\end{figure}
The FBM \cite{p26,phc10,hhp15} captures the fracture dynamics in composite
materials. In this model, a large number of parallel Hookean
springs or fibers are clamped between two horizontal platforms;
the upper one (rigid) helps hanging the bundle while the load hangs
from the lower one (Figure \ref{fig:FBM-model}). The springs or fibers are 
assumed to have
identical spring constants though their breaking strengths are
different. Once the load per fiber exceeds a fiber's  
own threshold, it fails and can not carry load any more. 
The load it carried is now shared by the
surviving fibers. If the lower platform deforms under loading, fibers 
closer to the just-failed fiber will absorb more of the load compared to 
those further away and this is called local-load-sharing (LLS) scheme. On the 
other hand, if the lower platform is rigid, the load is equally distributed to 
all the surviving fibers. This is called the equal-load-sharing (ELS) scheme. 
Obviously, for low values of the initial
load (per fiber), the successive failures of the fibers 
due to extra load sharing remain localized and though the
strain of the bundle (given by the identical strain of the
surviving fibers) grow with increasing load, the bundle as a
whole does not fail. Beyond a ``critical" value of the initial load, 
determined by the fiber strength distribution and the
load sharing mechanism (after each failures), the successive
failures become a global one and the bundle fails. Here,
the ``order" could be measured by the fraction of eventually
surviving fibers (after a ``relaxation time" required for
stabilization of the bundle), which decreases  as the load
on the bundle increases. Beyond the critical load, mentioned
above, the eventual damage size becomes global and the order
disappears (bundle fails). We will see in the following sections, as we
approach the critical load (either from higher or lower
load), the relaxation time (time or steps required since the load is applied 
until no further failure occurs or until the whole bundle collapses) 
diverges with ``universal" values
of the exponents (power) similar to the power laws in thermodynamic phase
transitions discussed in the previous section.

Long back, in $1926$, F. T. Peirce introduced the Fiber Bundle Model \cite{p26} 
to study 
the strength of 
cotton yarns in connection with textile engineering.  
Some static behavior of such a bundle (with equal load sharing by all the 
surviving fibers, following a failure) was discussed by Daniels in $1945$ 
 \cite{d45} and the model was brought to the attention of physicists  in $1989$ 
by Sornette \cite{s89}. 

\section{Equal Load Sharing FBM}

Let us consider a fiber bundle model having 
$N$ parallel fibers placed between two stiff clamps.  Each fiber responds
linearly with a force $f$ to an extention or stretch $\Delta$,
\begin{equation}
\label{eq1}
f=\kappa\Delta\;,
\end{equation}
where $\kappa$ is the spring constant.  We consider $\kappa=1$ for all 
fibers. 
Each fiber has a load threshold $x$ assigned to it.  If the stretch $\Delta$ 
exceeds this threshold, the fiber fails irreversibly.  In the 
equal-load-sharing (ELS) mode, the clamps are 
stiff and there is no non-uniform redistribution of loads among the surviving 
fibers, i.e., applied load is shared equally by the remaining intact fibers. 

\subsection{Fiber strength distributions and the cumulative distributions }
\begin{figure}[h]
\begin{center}
\includegraphics[width=6cm]{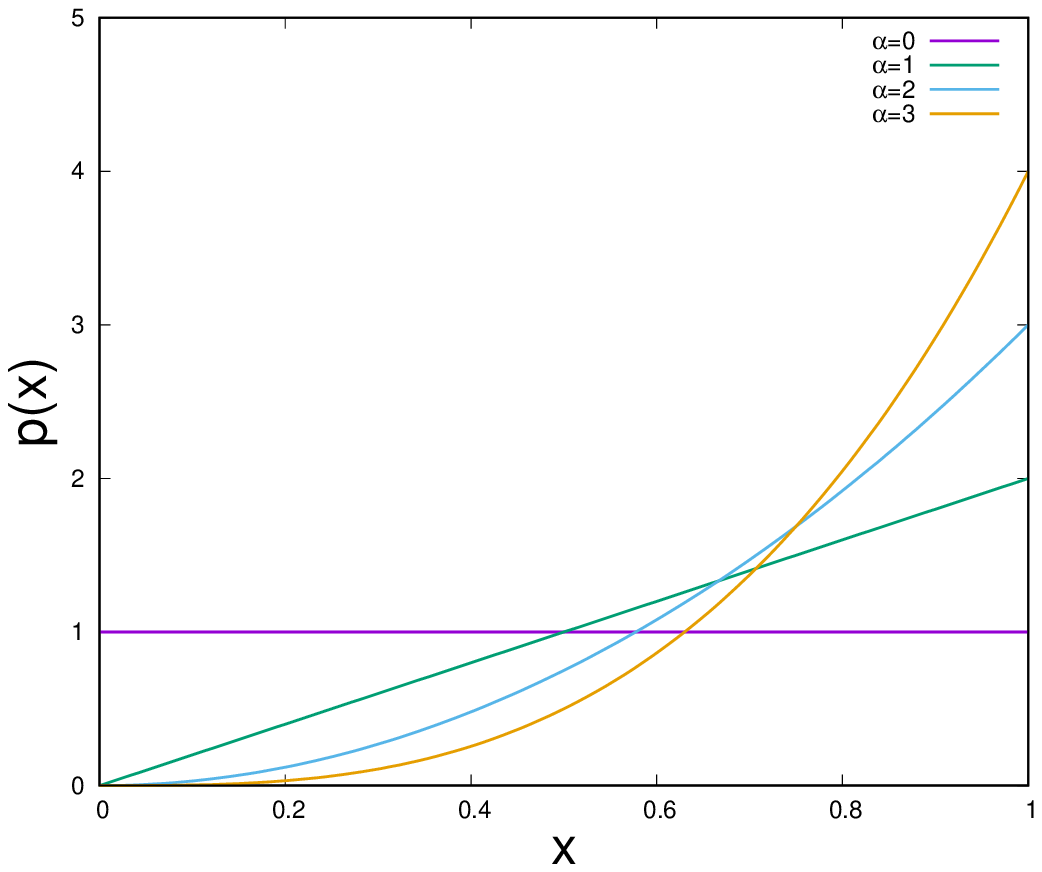}
\includegraphics[width=6cm]{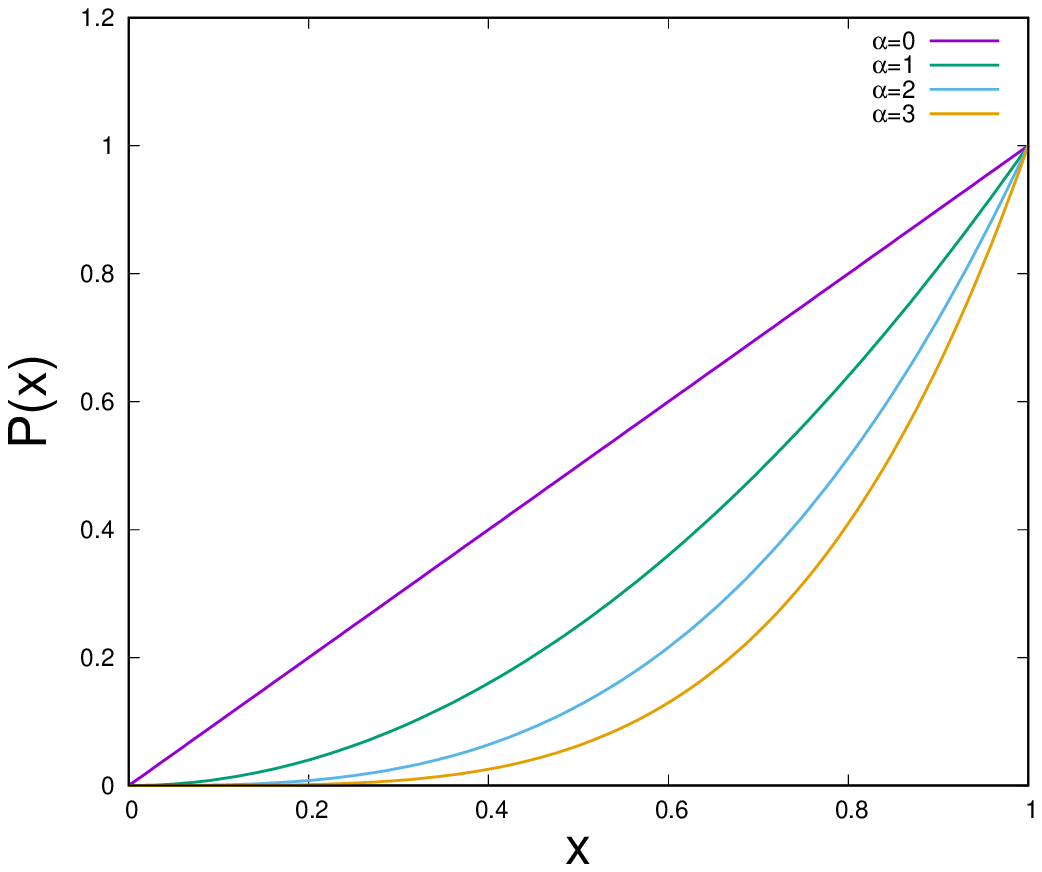}
\caption {\label{fig:Power-law-dists}
The power-law distributions of fiber thresholds and the corresponding 
cumulative distributions. The distribution reduces to an uniform distribution 
when power-law index $\alpha=0$.}
\end{center}
\end{figure}

\begin{figure}[h]
\begin{center}
\includegraphics[width=6cm]{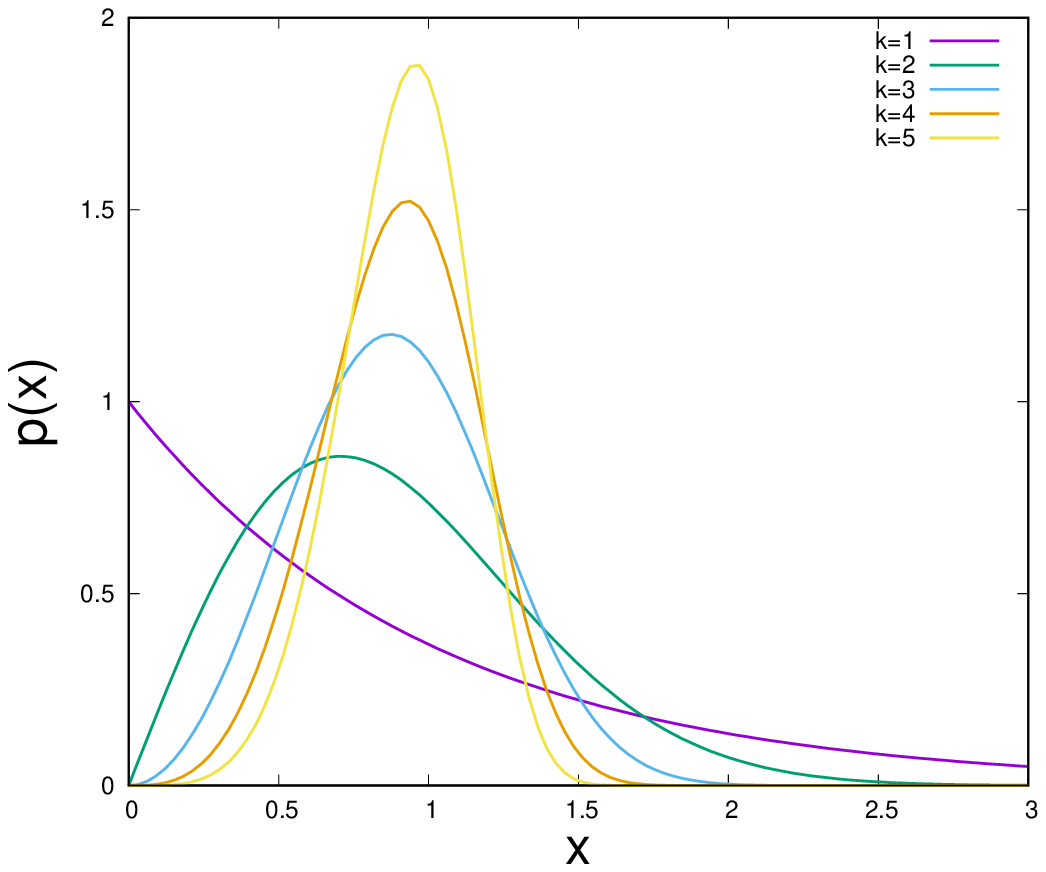}
\includegraphics[width=6cm]{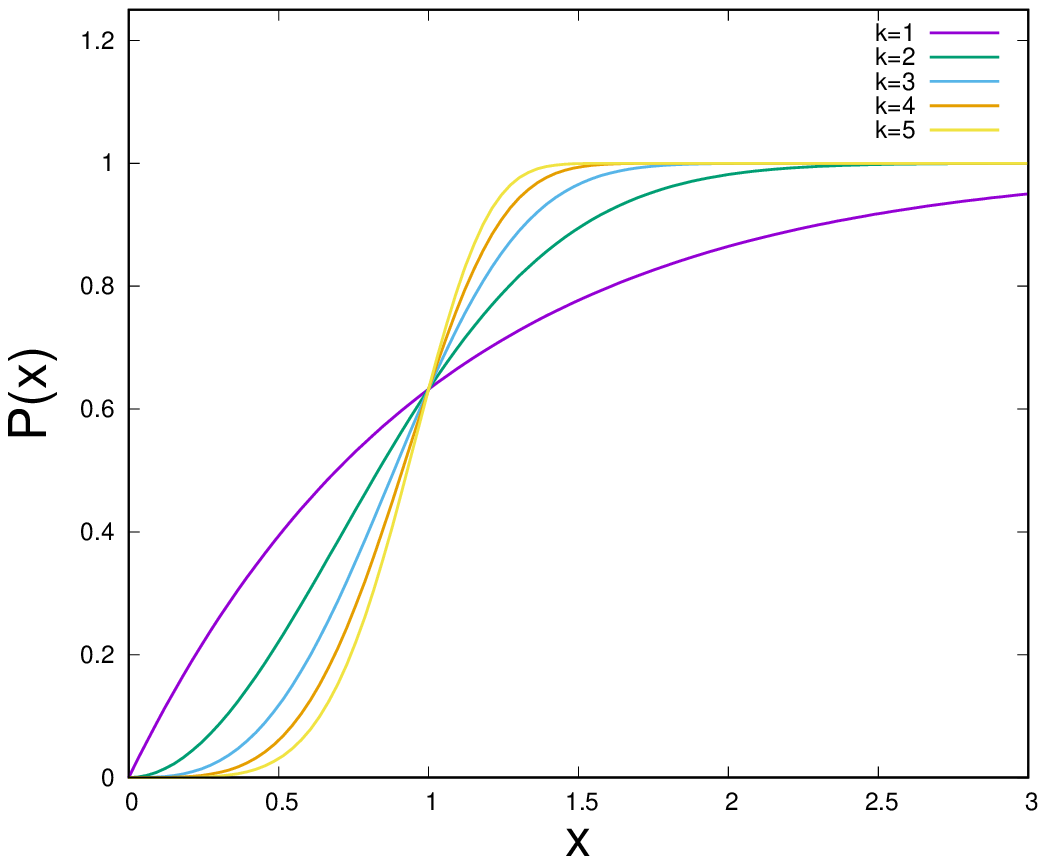}
\caption {\label{fig:Weibull-dists}
The  Weibull distributions of fiber thresholds and the corresponding 
cumulative distributions.}
\end{center}
\end{figure}

The fiber strength thresholds
are drawn from a probability density $p(x)$. The corresponding cumulative probability
is given by
\begin{equation}
\label{eq2}
P(y)=\int_0^y\  p(x) dx\;.
\end{equation}
The most studied threshold distributions are  Power-law type distributions and 
Weibull distributions (see Figures \ref{fig:Power-law-dists}, \ref{fig:Weibull-dists}). 

We consider a general power law type fiber threshold distributions within 
the range $(0,1)$,
\begin{equation}
\label{eq-norm}
p(x)\propto x^{\alpha}; \alpha\geq0.
\end{equation}
For normalization, we need to fulfil
\begin{equation}
\label{eq-norm-cond}
\int_0^1 p(x)dx=1.
\end{equation}
Therefore we get, from Eqns. (\ref{eq-norm}, \ref{eq-norm-cond}), the 
prefactor is $(1+\alpha)$:  
\begin{equation}
\label{eq-full}
p(x)=(1+\alpha)x^{\alpha}.
\end{equation}
The cumulative distribution takes the form
\begin{equation}
\label{eq-cumm}
P(x)= \int_0^x p(y)dy =x^{1+\alpha}.
\end{equation}
When the power-law index $\alpha=0$, the distribution reduces to an uniform 
distribution with 
\begin{equation}
\label{eq:uniform}
p(x)=1;   P(x)=x.
\end{equation}
In Figure (\ref{fig:Power-law-dists}) we present the probability distributions 
 and corresponding cumulative distributions for power-law type threshold 
distributions. 

On the other hand the cumulative
Weibull distribution has a form:
\begin{equation}
\label{eq:Weibull-cumm}
P(x)= 1-\exp(-x^k),
\end{equation}
where, $k$ is the shape parameter -sometimes called Weibull index. 
Therefore the probability distribution takes the form:
\begin{equation}
\label{eq:Weibull-prob}
p(x)=kx^{k-1}\exp(-x^k).
\end{equation}
In figure (\ref{fig:Weibull-dists}) we present the probability distributions 
 and corresponding cumulative distributions for Weibull threshold 
distributions. 

\subsection{The Load Curve and the Critical values}
\begin{figure}[t]
\begin{center}
\includegraphics[width=6cm]{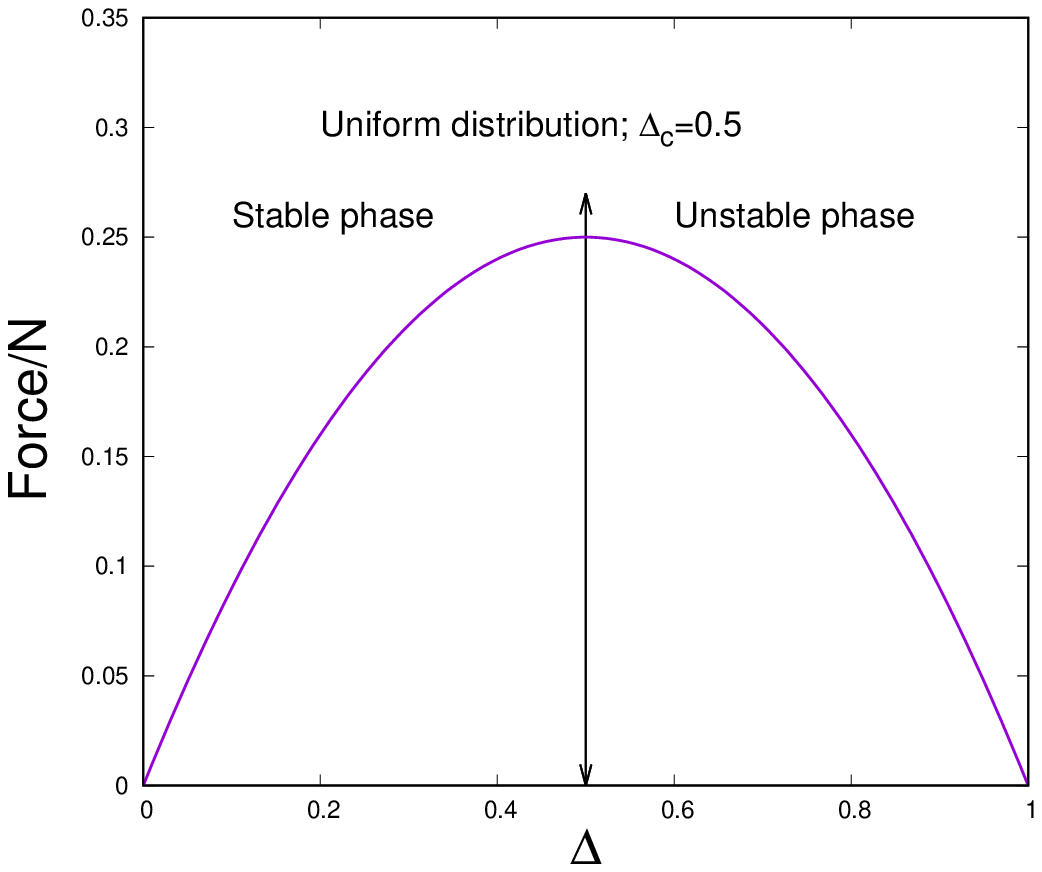}
\includegraphics[width=6cm]{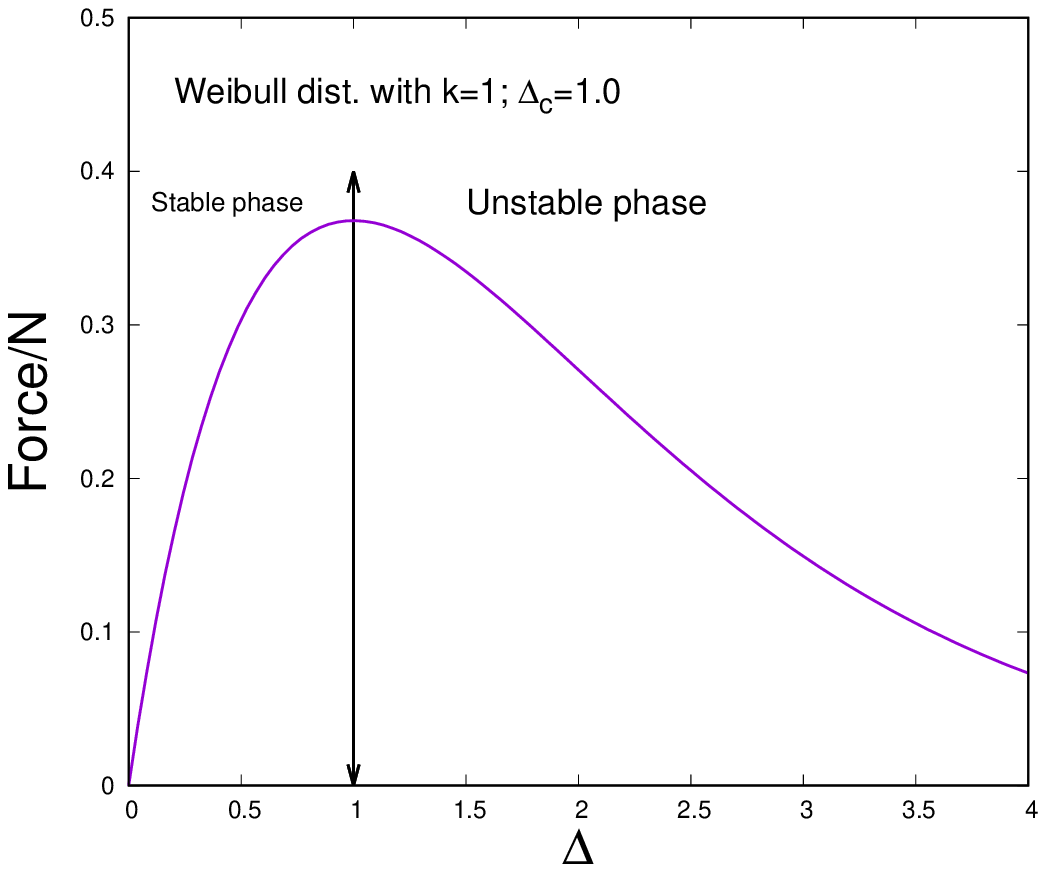}
\caption {\label{fig:phase-uniform-Weibull}
Normalized force vs. extention $\Delta$ for a fiber 
bundle with uniform and Weibull distributions of thresholds.}
\end{center}
\end{figure}
When the fiber bundle is loaded, the fibers fail according to their thresholds, the weaker
before the stronger.  We suppose that $N_f$ fibers have failed at a stretch or 
load 
$\Delta$. Then the fiber bundle supports a force
\begin{equation}
\label{eq3}
F=\kappa(N-N_f)\Delta =(N-N_f)\Delta,
\end{equation}
as the spring constant has been set equal to unity.  
This is a discrete picture and above equation is valid for any $N$ value, 
small or large.
When $N$ is very large, the force on the bundle at a stretch value $\Delta$ can 
be written as 
\begin{equation}
\label{eq:load-curve}
F=(N-N_f)\Delta=N(1-P(\Delta))\Delta.
\end{equation}
If we plot the normalized force ($F/N$) vs. strech value $\Delta$, we normally 
get a parabola like shape (Figure \ref{fig:phase-uniform-Weibull}).

\begin{figure}[h]
\begin{center}
\includegraphics[width=6cm]{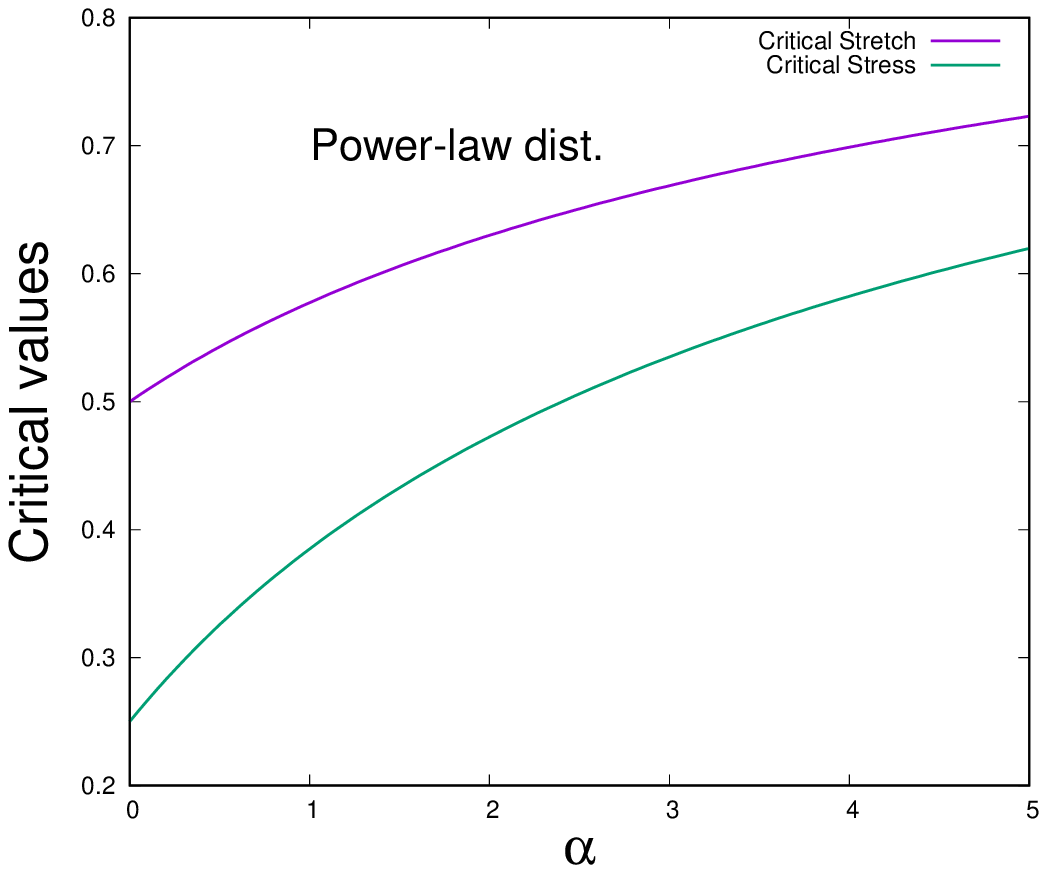}
\includegraphics[width=6cm]{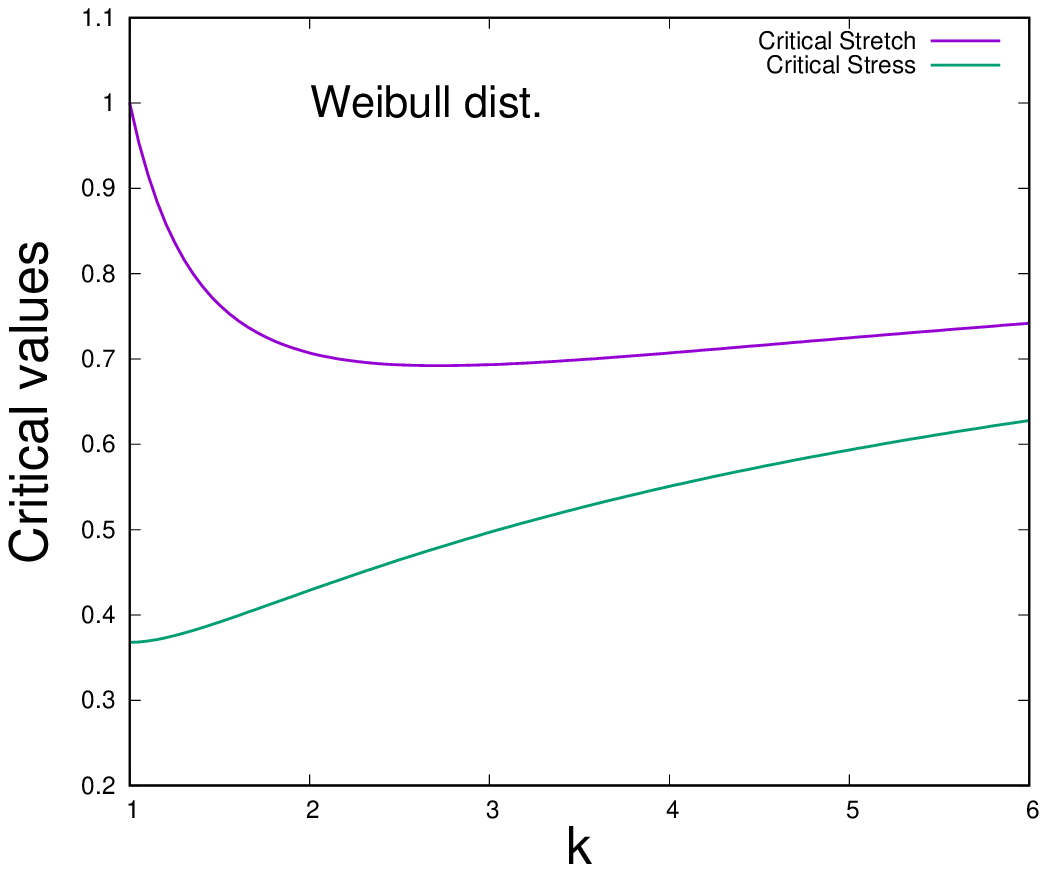}
\caption {\label{fig:critical-values}
Critical stretch and stress values vs. $\alpha$ (for power-law type distributions) and vs. $k$ for Weibull distributions of thresholds.}
\end{center}
\end{figure}

The force  has a maximum at a particular $\Delta$ value  ($\Delta_c$) and this 
is the maximum strength of the whole bundle. This point is often called the 
failure point or critical point of the system, beyond which the bundle 
collapses.  
Therefore we 
can say -there are two distinct phases of the system: stable phase for 
$0<\Delta\le\Delta_c$ and unstable phase for $\Delta > \Delta_c$.
Now, at the critical point, setting $dF(\Delta)/d\Delta =0$ we get
\begin{equation}
\label{eq13}
1-\Delta_c p(\Delta_c)-P(\Delta_c)=0.
\end{equation}
\subsubsection{General threshold distribution}
At the critical stretch $(\Delta_c)$, we recall Eq. (\ref{eq13})
 and putting the $p(\Delta_c)$ and $P(\Delta_c)$ values for a general 
 power-law type distribution, we get  
\begin{equation}
\label{eq-delta_c}
\Delta_c =\left(\frac{1}{2+\alpha}\right)^\frac{1}{1+\alpha}.
\end{equation}
What is the critical strength of the bundle?
If we put the $\Delta_c$ value in the force expression 
(Eq. \ref{eq:load-curve}), we get 
\begin{equation}
\label{eq-F_c-power}
\frac{F_c}{N} =(1+\alpha) \left(\frac{1}{2+\alpha}\right)^\frac{2+\alpha}{1+\alpha}.
\end{equation}
Inserting $\alpha=0$, we get 
\begin{equation}
\label{eq14}
\Delta_c=\frac{1}{2};\frac{F_c}{N}=\frac{1}{4}; 
\end{equation}
which are the critical stretch and force values for uniform distribution 
(Figures \ref{fig:phase-uniform-Weibull}, \ref{fig:critical-values}).  

\subsubsection{Weibull threshold distribution}
Let us move to a more general distribution of fiber thresholds, the Weibull 
distribution, which has been used widely in material science. 
As the force has a maximum at the failure point $\Delta_c$, recalling the 
expression (Eq. \ref{eq13}) and putting the Weibull $P(x)$, $p(x)$ values into 
it, we get
\begin{equation}
\label{eq-Weibull-Delta_c1}
\exp(-\Delta_c^k) - (\Delta_c k\Delta_c^{k-1}\exp(-\Delta_c^k)) = 0.
\end{equation}
From the above equation we can easily calculate the critical stretch value 
\begin{equation}
\label{eq-Weibull-Delta_c2}
\Delta_c =k^{-\frac{1}{k}};
\end{equation}
and the critical force value
\begin{equation}
\label{eq-F_c-Weibull}
\frac{F_c}{N} = k^{-\frac{1}{k}}\exp({-\frac{1}{k}}).
\end{equation}
For $k=1$, $\Delta_c=1.0$ and $\frac{F_c}{N}=\frac{1}{e}$ (Figures \ref{fig:phase-uniform-Weibull}, \ref{fig:critical-values}).
\subsection{Quasi-static loading vs. loading by discrete steps}
So far we have described the ELS model and derived the equlibrium 
Force-elongation or stress-strain relation. We did not say anything about the 
way of loading, i.e., how the load/force has been applied to the bundle ?

\begin{figure}[h]
\begin{center}
\includegraphics[width=8cm]{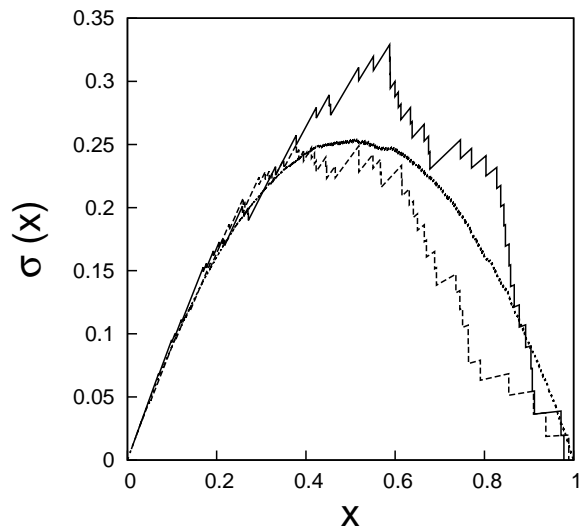}
\caption {\label{fig:force-fluctuation}
Two realizations for the force per fiber $F(x)/N=\sigma(x)$ as a function of 
the stretch $x$ for a bundle with $N=50$ having uniform distribution of fiber
 thresholds. For comparison, a realization with $N=5000$ is shown. Clearly, 
for large number of fibers, the fluctuations are tiny and the resulting 
force-stretch curve almost follows the parabolic average force expression 
$x(1-x)$.}
\end{center}
\end{figure}

Going back to the history of FBM we find that people discussed first the 
``weakest-link-failure" mode of loading. This is a very slow loading process 
that ensures the breaking of only the weakest element (among the intact fibers).
This is clearly a ``quasi-static" approach and noise or fluctuation in 
threshold distribution plays a major role (Figure \ref{fig:force-fluctuation}) 
in this type of loading process. 

However, a fiber bundle can be loaded in a different way.
If a {\it finite} external force or load is applied, all fibers that cannot 
withstand 
the applied stress, fail. The stress on the surviving fibers then increases, 
which 
drives further fibers to break, and so on. This iterative breaking process will
 go on until an equilibrium with some intact fibers (those can support 
the load) 
 is reached or the whole 
bundle collapses. We are now going to study the {\it average} behavior of 
such breaking 
processes for a bundle of large number of fibers following the formulations 
in the References \cite{pc01,bpc03,ph07,phc10,hhp15}.

\subsection{Loading by discrete steps: The recursive dynamics}
Let us assume that an external force $F$ is applied to the fiber bundle, with the applied stress denoted by
\begin{equation}
\sigma = F/N,
\label{eq:2.4.0.1}\end{equation}
the external load per fiber. We let $N_t$ be the average number of fibers that 
survive after $t$ steps in the stress redistribution process, with $N_0 = N$. 
We want to determine how $N_t$ decreases until the degradation process stops. 

At a stage during the breaking process when $N_t$ intact fibers remain, the effective stress  becomes
\begin{equation}
x_t = N\sigma /N_t.
\label{eq:2.4.0.2}\end{equation}
Thus 
\begin{equation}
N P(N\sigma/N_t)
\label{eq:2.4.0.3}\end{equation}
of fibers will have thresholds that cannot withstand the load. In the next 
step, therefore, number of intact fibers will be
\begin{equation}
 N_{t+1} = N - NP(N\sigma/N_t).
\label{eq:2.4.0.4}\end{equation}
Now we define the relative number of intact fibers as 
\begin{equation}
n_t = N_t/N,
\label{eq:2.4.0.5}\end{equation}
therfore, Eqn. (\ref{eq:2.4.0.4}) takes the form of a nonlinear recursion 
relation, 
\begin{equation}
n_{t+1} = 1 - P(\sigma/n_t), 
\label{eq:2.4.0.6}\end{equation}
with $\sigma$ as the control parameter and with $n_0=1$ as the start value.

We can also set up a recursion for $x_t$, the effective stress $\sigma/n_t$ after $t$ iterations:
\begin{equation}
x_{t+1} = \frac{\sigma}{1-P(x_t)},
\label{eq:2.4.0.7}\end{equation} 
with $x_0=\sigma$ as the initial value. Since by  (\ref{eq:2.4.0.2})
\begin{equation}
x_t=\sigma/n_t,
\label{eq:2.4.0.8}\end{equation}
the two recursion relations (\ref{eq:2.4.0.6} and \ref{eq:2.4.0.7}) 
may be mapped onto each other.

In general it is not possible to solve  nonlinear iterations like 
(\ref{eq:2.4.0.6}) or (\ref{eq:2.4.0.7}) analytically. The model with  
uniform ($\alpha=0$) and linerly increasing ($\alpha=1$) threshold 
distributions are  however, 
exceptions.

In nonlinear dynamics the character of an iteration is primarily determined by 
its {\it fixed points} (denoted by *). We are therefore interested in 
possible fixed points 
$n^*$  of (\ref{eq:2.4.0.6}), which satisfy
\begin{equation}
n^* = 1- P(\sigma/n^*).
\label{eq:2.4.1.1}\end{equation} 
Correspondingly, fixed points $x^*$ of the iteration (\ref{eq:2.4.0.7}) must satisfy 
\begin{equation}
x^* = \frac{\sigma}{1-P(x^*)},
\label{eq:2.4.1.2}\end{equation}
which may be written as 
\begin{equation}
F= N\sigma = N x^*(1-P(x^*)).
\label{eq:2.4.1.3}\end{equation}
This is precisely the relation (\ref{eq:load-curve}) between stress and strain. 
Therefore the equilibrium value of $x$, for a given external stress $\sigma$, 
is a fixed point. 

\subsection{Solution of the recursive dynamics and the Critical exponents}
Let us illustrate these general results by an example. 
We consider first a power-law type distribution (\ref{eq-full})
\begin{equation}
\label{eq-power-all}
p(x)=(1+\alpha)x^{\alpha}; P(x)=x^{\alpha+1}.
\end{equation}
The fixed-point equation  (\ref{eq:2.4.1.1}) takes the form
\begin{equation} 
(n^*)^{\alpha+2} -(n^*)^{\alpha+1} +\sigma^{\alpha+1} =0.
\label{eq-fixed-power}\end{equation}
If we set $\alpha=0$,  the threshold distribution reduces to an uniform 
threshold distribution  and the recursion relation becomes 
\begin{equation}
n_{t+1}= 1-\sigma/n_t.
\label{eq:recur-uni}
\end{equation}
Consequently the fixed point equation assumes 
a harmless form 
\begin{equation} 
(n^*)^2 -n^* +\sigma =0 ,
\label{eq:2.4.1.9}\end{equation}
with solution
\begin{equation}
n^* = {\textstyle \frac{1}{2}} \pm \left(\sigma_c-\sigma\right)^{1/2}.
\label{eq:2.4.1.10}\end{equation}
Here  $\sigma_c=1/4$, the critical value of the applied stress beyond which the bundle fails completely.  
In (\ref{eq:2.4.1.10}) the upper signs give $n^*>n_c$ which corresponds to 
stable fixed points. The -ve sign in (\ref{eq:2.4.1.10}) is unphysical as 
for $\sigma=0$ it gives $n^*=0$. 
From this solution we can easily derive order parameter, susceptibility and 
relaxation time behavior and their exponents.
\subsubsection{Order parameter}
From the fixed-point solution we  get at the critical point 
($\sigma =\sigma_c$)
\begin{equation}
n_c^* = \frac{1}{2}.
\label{en:nc-uniform}\end{equation}
Therefore we can present the fixed-point solution as
\begin{equation}
n^*(\sigma)-n_c^* \propto (\sigma_c-\sigma)^{\beta}, \hspace{8mm} \beta = {\textstyle \frac{1}{2}},
\label{eq:order-uni}\end{equation}
where $n^*(\sigma)-n_c^*$  behaves like an order parameter, i.e., it shows a 
transition from non-zero to zero value at the critical point $\sigma_c$.
\subsubsection{Susceptibility}
We can define the breakdown susceptibility as $\chi=-dn^*/d\sigma$. From the 
fixed-point solution we can write directly
\begin{equation}
\chi \propto (\sigma_c-\sigma)^{-\gamma}, \hspace{8mm} \gamma = {\textstyle \frac{1}{2}};
\label{eq:kappa-uni}\end{equation}
 which diverges at the critical point following a well-defined power law. This 
is another robust signature of a critical phenomenon.
\begin{figure}[h]
\begin{center}
\includegraphics[width=8cm]{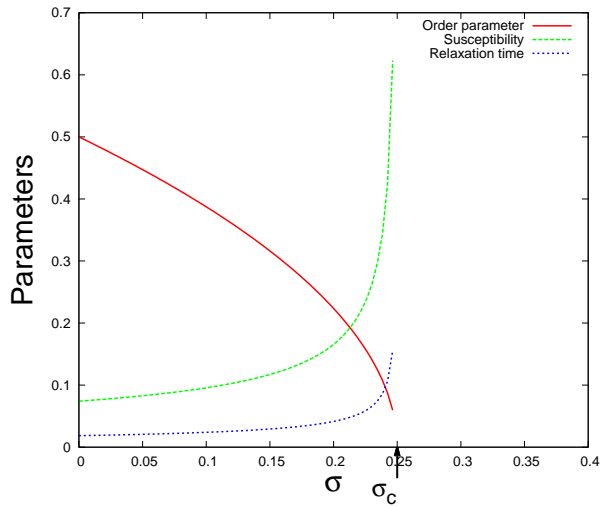}
\caption {\label{fig:parameters-uni}
Variation of order parameter, susceptibility and relaxation time vs. applied stress $\sigma$ for uniform distribution of fiber strengths.}
\end{center}
\end{figure}
\subsubsection{Relaxation time}
To track down the  approach very near a fixed point, we note that close to a stable fixed point the iterated quantity changes by tiny amounts, so that one may expand in the differences $n_t-n^*$. For the model with uniform distribution of the thresholds, the recursion relation (\ref{eq:2.4.0.6}), 
\begin{equation}
n_{t+1}= 1-\sigma/n_t,
\label{eq:2.4.1.12}
\end{equation}
 gives to linear order
\begin{equation}
n_{t+1}-n^* = \frac{\sigma}{n^*}-\frac{\sigma}{n_t}=\frac{\sigma}{n_tn^*}\;(n_t-n^*) \simeq \frac{\sigma}{n^{*2}} (n_t-n^*).
\label{eq:2.4.1.13}
\end{equation}
Thus the fixed point is approached monotonously with exponentially decreasing steps:
\begin{equation}
n_t-n^* \propto e^{-t/\tau},
\label{eq:2.4.1.14}\end{equation}
with a relaxation parameter
\begin{equation}
\tau = 1/\ln(n^{*2}/\sigma)= 1\Big/\ln\left[\left({\textstyle\frac{1}{2}}+\sqrt{{\textstyle\frac{1}{4}}-\sigma}\right)^2\Big/\sigma\right] .
\label{eq:2.4.1.15}\end{equation}
For the critical load, $\sigma = \sigma_c=\frac{1}{4}$, the argument of 
the logarithm is $1$, so that apparently $\tau$ is infinite. More precisely, 
for $\sigma \rightarrow \sigma_c$ 
\begin{equation}
\tau \simeq {\textstyle \frac{1}{4}}\;(\sigma_c-\sigma)^{-\theta}\hspace{5mm}\mbox{ with }\hspace{3mm} \theta = {\textstyle \frac{1}{2}}.
\label{eq:2.4.1.16}\end{equation}
 The divergence is a clear indication that the character of the breaking dynamics changes when the bundle goes critical.\\
\subsubsection{Critical slowing}
How does the system behave at the critical point ? If we put the critical $\sigma$ value in the recursive equation, we get
 
\begin{equation}
n_{t}-n_c^*\sim t^{-\phi}, \phi=1,
\label{eq:critical-slowing-uni}
\end{equation}
which implies the relaxation dynamics is critically slow exactly at the 
critical stress value. Critical-slowing is another known characteristic of a
critical phenomenon.
 
\subsection{Universality}
So far we obtained the dynamic critical behavior for the uniform 
distribution of the 
breaking thresholds, and the natural question is how general the results are. 
We can do a spot check on universality through considering a different 
distribution of fiber strengths.  By setting $\alpha=1$,  
the power-law type distribution reduces to a {\it linearly increasing 
distribution} on the
 interval $(0,1)$,
\begin{equation}
p(x) = \left\{\begin{array}{ll}
2x, & \hspace{5mm}0\leq x \leq 1\\
0& \hspace{5mm} x >1.
\end{array} \right.
\label{eq:2.4.3.1}\end{equation} 
For simplicity nondimensional variables are used.
By the force-elongation relationship the average total force per fiber,
\begin{equation}
F(x)/N = \left\{ \begin{array}{ll}
x(1-x^2)& \hspace{5mm} 0\leq x \leq 1\\
0       & \hspace{5mm} x>1,
\end{array}\right. ,
\label{eq:2.4.3.2}\end{equation}
shows that the critical point is 
\begin{equation}
x_c=\frac{1}{\sqrt{3}}, \hspace{8mm} \sigma_c= \frac{2}{3\sqrt{3}}. 
\label{eq:2.4.3.3}\end{equation}
In this case the recursion relation takes the form
\begin{equation}
n_{t+1}= 1-(\sigma/n_t)^2,
\label{eq:recur-inc}
\end{equation}
consequently the fixed-point equation reduces to 
\begin{equation}
(n^*)^{3} -(n^*)^{2} +\sigma^{2} =0.
\label{eq:fixed-inc}\end{equation}
a cubic equation in $n^*$. Therefore there exist three solutions of $n^*$ for 
each value of $\sigma$. 
For the critical load, $\sigma_c=2/3\sqrt{3}$, the only real and positive 
solution of (\ref{eq:fixed-inc}) is 
\begin{equation}
n^*_c = {\textstyle \frac{2}{3}}.
\label{eq:2.4.3.10}\end{equation}
One can show that for $\sigma <\sigma_c$ there will be an unstable fixed point with $n^*<n^*_c$, and a stable one with $n^*>n^*_c$. 

To find the number of intact fibers in the neighborhood of the critical point, we insert $n=\frac{2}{3}+(n-n_c)$ into (\ref{eq:recur-inc}), with the result
\begin{equation}
{\textstyle \frac{4}{27}} - (n-n_c)^2-(n-n_c)^3 = \sigma^2 = ({\textstyle \frac{2}{3\sqrt{3}}}+\sigma-\sigma_c)^2={\textstyle \frac{4}{27}}+{\textstyle \frac{4}{3\sqrt{3}}} (\sigma - \sigma_c) + (\sigma-\sigma_c)^2.
\label{eq:2.4.3.11}\end{equation}
To leading order we have
\begin{equation}
(n-n_c)^2 = {\textstyle\frac{4}{3\sqrt{3}}}(\sigma_c-\sigma).
\label{eq:2.4.3.12}\end{equation}
\subsubsection{Order parameter}
Hence for $\sigma \leq \sigma_c$ the order parameter behaves as
\begin{equation}
n(\sigma)-n_c \propto (\sigma_c-\sigma)^{\beta}, \hspace{8mm} \beta = {\textstyle \frac{1}{2}},
\label{eq:order-linear}\end{equation}
in accordance with (\ref{eq:order-uni}). 
\subsubsection{Susceptibility}
The breakdown susceptibility $\chi=-dn/d\sigma$ will therefore have the same 
critical behavior,
\begin{equation}
\chi \propto (\sigma_c-\sigma)^{-\gamma}, \hspace{8mm} \gamma={\textstyle \frac{1}{2}}
\label{eq:kappa-linear}\end{equation}
as for the model with a uniform distribution of fiber strengths.
\begin{figure}[h]
\begin{center}
\includegraphics[width=8cm]{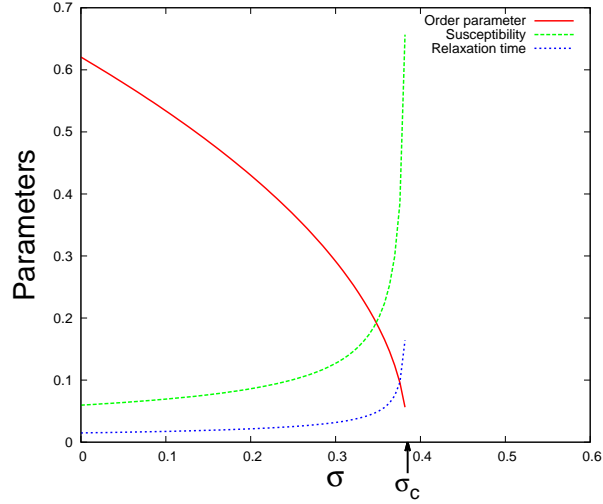}
\caption {\label{fig:parameters-inc}
Variation of order parameter, susceptibility and relaxation time vs. applied 
stress $\sigma$ for linearly increasing fiber strength distribution.}
\end{center}
\end{figure}

\subsubsection{Relaxation time}
Let us also investigate how the stable fixed point is approached.
From (\ref{eq:recur-inc}) we find
\begin{equation}
n_{t+1}-n^* = \frac{\sigma^2}{n^{*2}}-\frac{\sigma^2}{n_t^2} = \frac{\sigma^2}{n^{*2}n_t^2}\;(n_t^2-n^{*2} )\simeq (n_t-n^*)\;\frac{2\sigma^2}{n^{*3}}
\label{eq:2.4.3.15}\end{equation}
near the fixed point. Hence the approach is exponential, 
\begin{equation}
n_t-n^* \propto e^{-t/\tau} \hspace{8mm} \mbox{with } \hspace{4mm}\tau = \frac{1}{\ln (n^{*3}/2\sigma^2)}.
\label{eq:2.4.3.16}\end{equation}
At the critical point, where $n_c^*=2/3$ and $\sigma_c=2/3\sqrt{3}$,  the argument of the logarithm equals $1$, so that $\tau$ diverges when the critical state is approached. The divergence is easily seen to be of the same form, 
\begin{equation}
\tau \propto (\sigma_c-\sigma)^{-\theta}, \theta={\textstyle \frac{1}{2}},
\end{equation}
as for the model with a uniform threshold distribution, 
equation (\ref{eq:2.4.1.16}).

\subsubsection{Critical slowing}

To find the correct behavior of the distance to the critical point, 
$\Delta n_t=n_t-n_c =n_t-2/3$, {\it at criticality}, we use the iteration 
(\ref{eq:recur-inc}) with $\sigma=\sigma_c$,
\begin{equation}
n_{t+1} = 1-\frac{4/27}{n_t^2}, \hspace{8mm}\mbox{ or }\hspace{4mm} \Delta n_{t+1} = \frac{1}{3} -
\frac{4/27}{(\frac{2}{3}+\Delta n_t)^2}.
\label{eq:2.4.3.17}\end{equation} 
Near the fixed point, the deviation $\Delta n_t =n_t-n^*$ is small. An expansion to second order in  $\Delta n_t$ yields
\begin{equation}
\Delta n_{t+1} = \Delta n_t - {\textstyle \frac{9}{4}}\Delta n_t^2,
\label{eq:2.4.3.18}\end{equation}
which is satisfied by
\begin{equation}
\Delta n_t = \frac{4}{9t} + {\cal O}(t^{-2}).
\label{eq:2.4.3.19}\end{equation}
The slow critical relaxation towards the fixed point, 
\begin{equation}
n_t-n_c\propto t^{-\phi}, \phi=1,
\end{equation}
for large $t$,  is the same as for the uniform threshold distribution, formula (\ref{eq:critical-slowing-uni}). 

In conclusion, we have found that the model with a linearly increasing 
distribution of the fiber strengths possesses the same critical power laws as 
the model with a uniform distribution. This suggests that the critical properties are universal.

\subsection{Graphical solutions of the recursive dynamics}
\begin{figure}[t]
\begin{center}
\includegraphics[width=7cm]{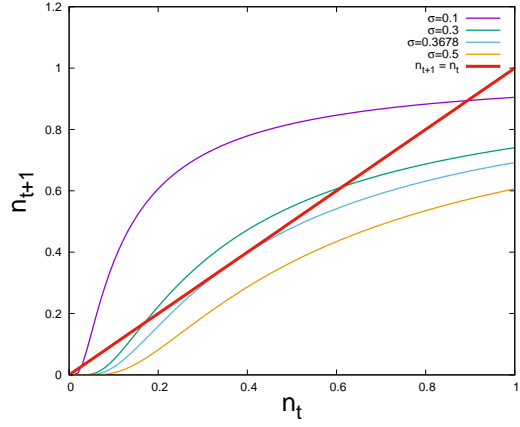}
\caption {\label{fig:graphical-Weibull}
Graphical solution of the ELS recursive dynamics for Weibull distribution.}
\end{center}
\end{figure}
Even if the recursive dynamics can not be solved for each and every fiber 
thresold distributions, through a graphical solution scheme one can always 
reach the critical points. The trick is to plot $n_{t+1}$ vs. $n_t$ and check
where this plot touches the fixed-point line $n_{t+1}=n_t$.
For example, let us consider a Weibull distribution (with $k=1$) of fiber 
thresholds (\ref{eq:Weibull-cumm}), having cumulative distribution
\begin{equation}
\label{eq:Weibull-k1}
P(x)= 1-\exp(-x).
\end{equation}
When an external stress $\sigma$ is applied, the recursion relation can be written as:
\begin{equation}
n_{t+1}= 1-[1-\exp(-\sigma/n_t]=\exp(-\sigma/n_t).
\label{eq:recur-Weibull}
\end{equation}
Now  we plot $n_{t+1}$  vs. $n_t$ following Eqn. \ref{eq:recur-Weibull} for 
several $\sigma$ values (see Figure \ref{fig:graphical-Weibull}).  
It is clear that at a particular $\sigma$ value, the $n_{t+1}$ vs. $n_t$ curve 
touches the $n_{t+1}=n_t$ straight line at a single point and this particular 
$\sigma$ 
value 
is the critical stress value $\sigma_c $ for this model.  

\subsection{Approach to the critical point}
It is important to find out how the system is approaching the critical point 
(failure point) from below (pre-critical) and above (post-critical) stress 
values. 

\begin{figure}[t]
\begin{center}
\includegraphics[width=7cm]{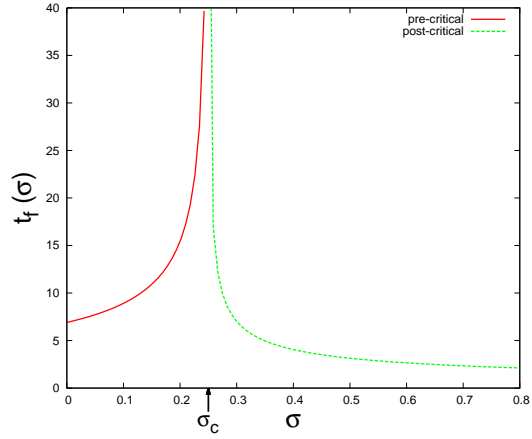}
\caption {\label{fig:approach}
Approach to critical point from below (pre-critical state) and from above 
(post-critical state) confirms two-sided critical divergence of the recursive 
step values.}
\end{center}
\end{figure}
For uniform fiber strength distribution when the external load approaches the 
critical load $\sigma_c=1/4$ from a 
higher value, i.e., in the post-critical region, the number 
of necessary iterations increases as one approaches critical point. 
Near criticality, number of iterations has a square-root divergence \cite{ph07}:
\begin{equation}
t_f \simeq {\textstyle \frac{1}{2}}\pi (\sigma - \sigma_c)^{-1/2}. 
\label{eq:2.4.4.15}\end{equation}
Similarly, in the pre-critical region, when the external load approaches the 
critical load $\sigma_c=1/4$ from a 
lower value, near the critical point the number of iterations has again a 
square root divergence \cite{ph07} (for uniform distribution), 
\begin{equation}
t_f = {\textstyle \frac{1}{4}}\;\ln (N)\,(\sigma_c-\sigma)^{-1/2},
\label{eq:2.4.5.15}\end{equation}
with a system-size-dependent amplitude.

We therefore find that in FBM, there exists a two-sided critical divergence 
behavior (Figure \ref{fig:approach}) in terms of the number of iteration steps 
needed to reach critical 
point from below (pre-critical) and above (post-critical) the critical point.
Detailed derivation of the above divergences in $t_f$ are given in the 
Appendix.

\subsection{Percolation in Equal Load Sharing FBM}
We will now discuss the connection between 
the Equal Load Sharing
Fiber Bundle Model and the standard percolation model.

In the ELS mode, when the fiber bundle is loaded, the fibers fail according to 
their thresholds, the weaker
before the stronger.
At a load or stretch $\Delta$, the fiber bundle supports a force
\begin{equation}
\label{eq3-v2}
F=N\left[1-P(\Delta)\right]\Delta, 
\end{equation}
where spring constant ($\kappa$) has been set to unity. 
Clearly, $P(\Delta)$ is the fraction of failed fibers at a load 
(extension) $\Delta$.
We know that there is a certain $\Delta=\Delta_c$
 value beyond which catastrophic failure occurs and the system collapses
completely. We are particularly interested to know whether  
 the cluster of broken fibers percolates in the stable phase (before the failure point is reached) or not?
To answer this question we need to calculate  $P(\Delta_c)$.
From our analysis in section IV.B, we recall that the force has a parabolic 
maximum at the failure point $\Delta_c$, where the following relation is valid: 
\begin{equation}
\label{eq13-v2}
1-\Delta_c p(\Delta_c)-P(\Delta_c)=0.
\end{equation}
Therefore
\begin{equation}
\label{eq-meso1}
P(\Delta_c) = 1-\Delta_c p(\Delta_c).
\end{equation}

\subsubsection{General threshold distribution}
We consider a general power law type fiber threshold distributions within
the range $(0,1)$,
\begin{equation}
\label{eq-full-2}
p(x)=(1+\alpha)x^{\alpha}; P(x)= x^{1+\alpha}.
\end{equation}
Putting the $p(\Delta_c)$ and $P(\Delta_c)$ values in eq. (\ref{eq13-v2} ) 
we get
\begin{equation}
\label{eq-delta_c}
\Delta_c =\left(\frac{1}{2+\alpha}\right)^\frac{1}{1+\alpha}.
\end{equation}

From the above relations, we can easily calculate the fraction of failed fibers at the failure point
\begin{equation}
\label{eq-meso2}
P(\Delta_c) = \frac{1}{2+\alpha}\leq\frac{1}{2}.
\end{equation}

\begin{figure}[t]
\begin{center}
\includegraphics[width=8cm]{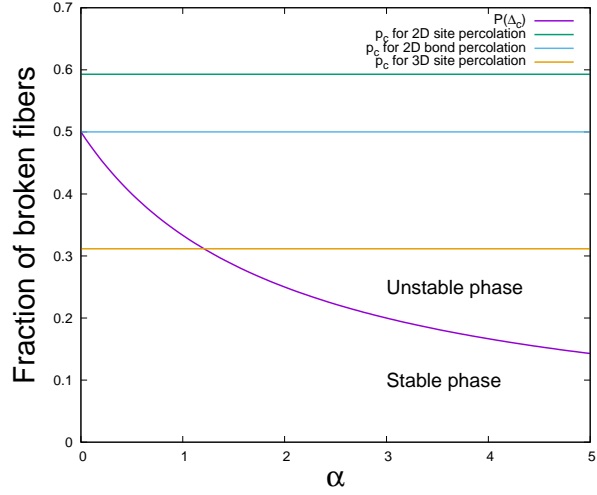}
\caption {\label{fig:perco-powerlaw}
Fraction of broken fibers vs. power law index $\alpha$. The curved line is $P(\Delta_c)$, which
seperates the stable and unstable phases of the FBM. Straight lines are the percolation
threshold values.}
\end{center}
\end{figure}

It is obvious that if $P(\Delta_c)<p_c$, ($p_c$ is the percolation threshold) 
largest cluster of broken fibers does not percolate until the failure point 
itself is reached.
Therefore in case of power-law type distribution of thresholds, we do not see 
percolation of broken fibers in 2D until the system enters into unstable phase. In three dimensional bond-percolation problem the situation is different 
-as long as 
 $\alpha<1.25$  cluster of brken fibers percolates in the stable phase  
(Figure (\ref{fig:perco-powerlaw})).

\subsubsection{Analaysis for Weibull threshold distribution}
Let us move to a more general distribution of fiber thresholds, the Weibull
distribution. The cumulative
Weibull distribution has a form:

\begin{equation}
\label{eq-Weibull-cumm}
P(x)= 1-\exp(-x^k),
\end{equation}
where, $k$ is the shape parameter.
Therefore the probability distribution takes the form:
\begin{equation}
\label{eq-Weibull-prob}
p(x)=kx^{k-1}\exp(-x^k).
\end{equation}
As the force has a maximum at the failure point $\Delta_c$, recalling the
expression (Eq. \ref{eq13}) and putting the $P(x)$, $p(x)$ values into it,
we get
\begin{equation}
\label{eq-Weibull-Delta_c1}
\exp(-\Delta_c^k) - (\Delta_c k\Delta_c^{k-1}\exp(-\Delta_c^k) = 0.
\end{equation}
From the above equation we can easily calculate the critical extention value
as
\begin{equation}
\label{eq-Weibull-Delta_c2}
\Delta_c =k^{-1/k}.
\end{equation}
The fraction of fibers at the failure point is
\begin{equation}
\label{eq-Weibull-meso}
P(\Delta_c)= 1-\exp(-\frac{1}{k}).
\end{equation}

\begin{figure}[t]
\begin{center}
\includegraphics[width=8cm]{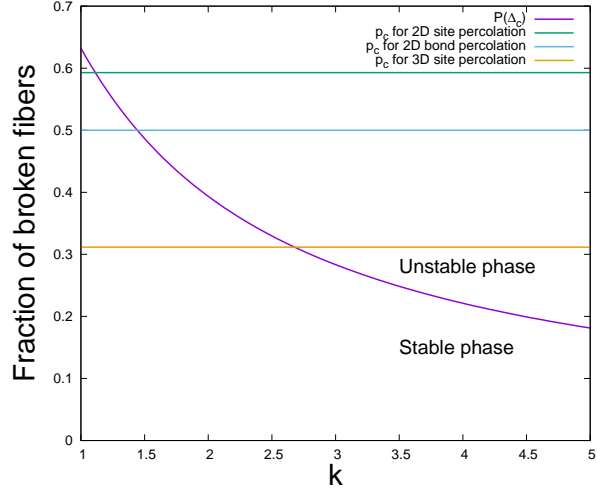}
\caption {\label{fig:perco-Weibull}
Fraction of broken fibers vs. Weibull index $k$. The curved line is $P(\Delta_c)$, which
seperates the stable and unstable phases of the FBM. Straight lines are the percolation
threshold values.}
\end{center}
\end{figure}

In case of Weibull distribution of fiber thresholds,
the clusters of broken fibers can percolate before the failure point is reached 
(for two and three dimensional site and bond 
percolation scenarios) when $k$ value remains within a certain window  
(Figure (\ref{fig:perco-Weibull})).

\section{Some related works on FBM}
In this section we would like to bring attention to some related works on 
FBM which, we believe,
may be regarded as essential reading in this field. 
If we follow quasi-static loading, the ELS model  produces avalanches 
(successive failure of fibers at a fixed external load) of 
different sizes during the entire fiber-failure process. The statistics of 
such avalanches were analysed by Hemmer and Hansen in their seminal paper 
in $1992$ \cite{hh92}. They found that the avalanches follow universal 
power law with 
exponent $-5/2$ for a mild restriction on the threshold distribution such that 
the load-curve (force vs. elongation) has single maximum. Later, Pradhan, Hansen and Hemmer showed that the exponent of avalnche distribution crosses over from 
$-5/2$ to $-3/2$ if we collect only the avalanches near the critical (failure) 
point \cite{phh05}. Also, Divakaran and Dutta studied \cite{dd07} the effect of 
discontinuity in the threshold distribution and obtained similar 
crossover behavior of 
avalanche distributions in ELS 
models.           

In $1991$, Harlow and Phoenix first introduced the Local-Load-Sharing (LLS) 
model \cite{hp91} with 
a simple breaking rule: when a fiber fails, the load it carried is shared 
by the nearest surviving fibers. The localized load-redistribution mechanism 
in LLS scheme makes the model very different from the ELS one. In one dimension,
 the 
critical strength of the LLS bundle shows a typical system size dependence 
\cite{gip93,phc10} 
: $\sigma_c(N)\sim 1/\ln(N)$, where $N$ is the total number of fibers in the 
chain. However, recent studies by NTNU group have established that in 
higher 
dimensions, memory independent LLS model shows non-zero critical 
strength \cite{skh15}.  
Biswas and Chakrabarti have studied the self-organized dynamics \cite{bc13}
 in LLS models 
by modifying the load-redistribution rule a bit, where the  steadily 
increasing external 
load is applied at a central point of the system. The redistributed 
load always remains localized along the steadily growing boundary of the 
broken patch and dynamic 
self-organization sets in. 
  
There have been some attempt to bridge the gap between ELS and LLS models 
by introducing some intermediate load-sharing rules. Hidalgo, Kun and Herrmann 
proposed a model \cite{hkh02} where the load that was carried by a broken fiber is 
redistributed to the surviving fibers following a decaying power law in the 
distance 
from the broken
fiber. In the same line, Pradhan, Chakrabarti and Hansen introduced a 
mixed-mode 
model \cite{pch05} where the ELS and LLS schemes are mixed together: When a fiber fails 
a fraction $g$ of the load it carried is distributed according to the LLS 
rule (to the fibers at the edge of the hole containg the broken fiber) and 
the rest 
$1-g$ fraction of the load is distributed to all the 
surviving fibers. Clearly, for $g=0$, the model reduces to a pure ELS model 
and for $g=1$, it is nothing but the LLS model. They found an interesting 
result that the mixed mode model crosses over 
from ELS to LLS behavior at $g \simeq 0.8$. 

Roy and Ray tried to see another type of critical behavior \cite{rr15} in 
ELS model in 
terms of brittle to quasi-brittle transition as a function of the width of 
the threshold distribution. Their claim is that at  (and below) a 
critical width value 
($\delta_c$), breaking of the weakest fiber leads to complete failure of 
the bundle and at $\delta_c$, relaxation time diverges obeying 
finite-
size scaling law: $\tau \sim N^{\beta}(|\delta-\delta_c|N^{\alpha})$ with 
$\alpha=\beta=1/3$.     

In section IV we have presented a mean-field treatment of the relaxation 
behavior of ELS models when the model is loaded by a discrete step. However,
one can expect finite-size dependence of the relaxation behavior at or around 
the critical stress values when the system size is  not large enough. 
Roy, Kundu and Manna have done extensive numerical studies \cite{rkm13} 
on the finite-size 
scaling forms of the relaxation time as a function of the deviation of stress 
values 
from the critical stress for the ELS model.
In Ref \cite{bs15} Biswas and Sen 
investigated, mostly analytically, the maximum strength and corresponding 
redistribution schemes
for sudden and quasistatic loading on FBM. The universality class associated
with the phase transition from partial to total failure (by increasing the load)
was found to be dependent on the redistribution mechanism.
 
The FBM has also been applied to traffic-jam modelling \cite{c06}, 
power-grid failure 
modelling \cite{pss14}, earthquake modelling, \cite{s92} etc. For a 
recent discussion on self-organized criticalities in FBM, for studying 
the propagation of crack front in heterogeneous solids, see \cite{pp18}.        
An elegant Renormalization 
Group Procedure 
in equal-load-sharing FBM analysis has been introduced in a very recent article \cite{phr18}. 
       
\section{Discussions}
The FBM is an extremely elegant model for studying the collective 
dynamical failure in inhomogeneus materials. The ELS scheme (ensured by the 
absolute rigidity of the platforms) of equally sharing of the extra load by 
the surviving fibers, after an individual fiber failure, allows often some 
precise analysis of the dynamics. With simple (uniform) fiber breaking 
threshold distribution, we have demonstrated here the ``critical behavior" 
and its related features like universality. The model here fits simple 
common sense, yet the collective failure dynamics in the bundle, its 
critical behavior, are extremely intriguing. Hope, the uninitiated 
readers can appreciate the excitement. 

The same model with realistic fiber threshold distribution (like Weibull 
distribution or Gumble distribution) may not always allow such analytic 
studies, but their numerical analysis (often with realistic LLS scheme for 
load sharing due to local deformations of the platforms with finite rigidity) 
can take one to the frontiers of civil engineering applications, as 
had been practiced by the professional engineers and architects. 

It may be noted, there is hardly any other model in science and technology,
 where some 
simplifications allow intriguing progress analytically in basic science, 
yet with some realistic ingredients added to the same model, it takes one 
to the 
forefront of engineering applications.  

\section{Acknowledgments}
The authors thank Alex Hansen 
for interesting discussions.  This work was partly supported by the
Research Council of Norway through its Centers of Excellence funding
scheme, project number 262644. BKC is grateful to J. C. Bose Fellowship Grant 
for support.
\section{Appendix}
\appendix {}
\section{Exact solutions for pre and post-critical relaxation}

The iterative breaking process considered in section IV ends with one of two 
possible end results. Either the whole bundle breaks down, or an equilibrium 
situation with a finite number of intact fibers is reached. The final fate 
depends on whether the external stress $\sigma$ on the bundle is postcritical  
($\sigma > \sigma_c$), precritical ($\sigma<\sigma_c$), or critical 
$\sigma=\sigma_c$.
We now investigate the total number $t_f$ of iterative steps $t$ necessary to 
reach the final state, and start with the {\it postcritical} situation 
following the formulations in References \cite{ph07,hhp15}.

\subsection{Postcritical relaxation}

For uniform threshold distribution (\ref{eq:uniform}) we can explicitly and 
exactly follow the path of iteration. We introduce a measure $\epsilon$ of 
the deviation from critical value by
\begin{equation}
\epsilon = \sigma -\sigma_c = \sigma - {\textstyle \frac{1}{4}},
\label{eq:2.4.4.1}
\end{equation}
where $\epsilon$ is positive.
The basic iteration formula (\ref{eq:2.4.0.6}) is in this case 
\begin{equation}
n_{t+1}= 1-\sigma/n_t. 
\label{eq:2.4.4.2}
\end{equation}
The fraction $n_t$ of intact fibers will decrease under the iteration, and we see from (\ref{eq:2.4.4.2}) that if $n_t$ reaches the value $\sigma$ or a smaller value, the next iteration yields $n_{t+1}=0$ or a negative value, i.e.\ complete bundle breakdown.  We wish to find  how many iterations, $t_f$, is needed to reach this stage.

For that purpose we solve the nonlinear iteration (\ref{eq:2.4.4.2}) by converting it into a linear iteration by means of two transformations.     
From (\ref{eq:2.4.4.2}), we can write 
\begin{equation}
n_{t+1} n_t = n_t-\sigma = n_t-{\textstyle \frac{1}{4}}-\epsilon.
\label{eq:2.4.4.3}\end{equation}
We introduce first
\begin{equation}
n_t = {\textstyle \frac{1}{2}}-y_t \sqrt{\epsilon},
\label{eq:2.4.4.4}\end{equation}
with the result
\begin{equation}
2\sqrt{\epsilon}=\frac{y_{t+1}-y_t}{1+y_{t+1}y_t}. 
\label{eq:2.4.4.5}\end{equation}
As a second transformation we put 
\begin{equation}
y_t=\tan v_t,
\label{eq:2.4.4.6}\end{equation}
with the result
\begin{equation}
2\sqrt{\epsilon} = \frac{\tan v_{t+1}-\tan v_t}{1+\tan v_{t+1}\;\tan v_t}= \tan (v_{t+1}-v_t).
\label{eq:2.4.4.7}\end{equation}
Hence we have now obtained the linear iteration
\begin{equation}
v_{t+1}-v_t = \tan^{-1} (2\sqrt{\epsilon}),
\label{eq:2.4.4.8}\end{equation}
with solution
\begin{equation}
v_t = v_0 + t \tan^{-1} (2\sqrt {\epsilon}).
\label{eq:2.4.4.9}\end{equation}

The iteration starts with all fibers intact, i.e.\ $n_0=1$, which by (\ref{eq:2.4.4.4}) and (\ref{eq:2.4.4.6}) corresponds to $y_0=-1/2\sqrt{\epsilon}$ and $v_0=-\tan^{-1} (1/2\sqrt{\epsilon})$. With the constant in (\ref{eq:2.4.4.9})  now 
determined, we can express the solution in terms of the original variable:
\begin{equation}
n_t = {\textstyle \frac{1}{2}}-\sqrt{\epsilon} \tan\left[-\tan^{-1}(1/2\sqrt{\epsilon})+t\; \tan^{-1}(2\sqrt{\epsilon})\right].
\label{eq:2.4.4.10}\end{equation}
We saw above that when $n_t$ obtains a value in the interval $(0,\sigma)$, the next iteration gives 
$n_{t+1}\leq 0$, complete bundle failure. $n_t$ reaching the smallest value $0$ gives
 an {\it upper} bound $t_f^u$ for the number of iterations. We find
\begin{equation}
t_f^u (\sigma)= 1+\frac{2\tan^{-1}(1/2\sqrt{\epsilon})}{\tan^{-1}(2\sqrt{\epsilon)}}.
\label{eq:2.4.4.11}\end{equation} 
And $n_t$ reaching the value $\sigma \equiv {\textstyle \frac{1}{2}}+\epsilon $ gives a {\it lower} bound:
\begin{equation}
t_f^l (\sigma) = 1+ \frac{\tan^{-1}\left[({\textstyle \frac{1}{4}}-\epsilon)/\sqrt{\epsilon}\right]+\tan^{-1}(1/2\sqrt{\epsilon})}{\tan^{-1}(2\sqrt{\epsilon})}.
\label{eq:2.4.4.12}\end{equation}

\begin{figure}[h]
\begin{center}
\includegraphics[width=8cm]{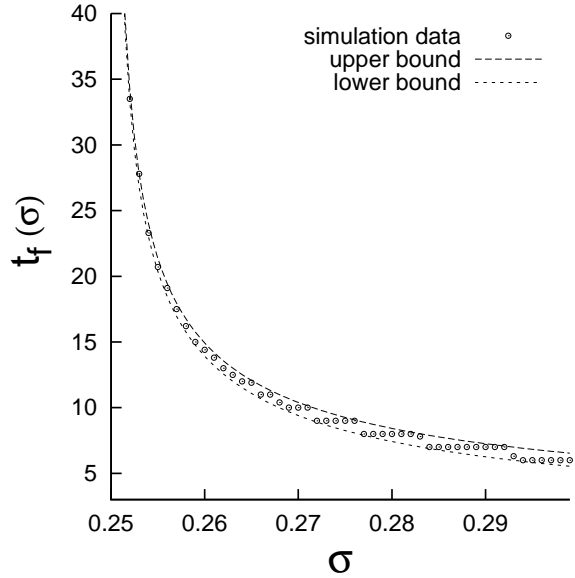}
\caption {\label{fig:post-crit}
Post-critical relaxation: Numerical data are for a bundle with $N=10^6$ fibers 
having uniform threshold distribution and averages are taken over 
$10^5$ samples.} 
\end{center}
\end{figure}
The 
upper and lower bounds (\ref{eq:2.4.4.11}) and (\ref{eq:2.4.4.12}) nicely 
embrace the simulation results (Figure (\ref{fig:post-crit})). 
When the external load is large, just a few iterations suffice to induce completely bundle breakdown.
On the other hand, when the external load approaches the critical load $\sigma_c=1/4$, the number of necessary iterations becomes vary large. 
Near criticality ($\epsilon \rightarrow 0$) both the upper and lower bounds, (\ref{eq:2.4.4.11}) and (\ref{eq:2.4.4.12}), have a square-root divergence:
\begin{equation}
t_f \simeq {\textstyle \frac{1}{2}}\pi (\sigma - \sigma_c)^{-1/2},
\label{eq:2.4.4.15}\end{equation}
to dominating order for small $\epsilon$. 

\subsection{Precritical relaxation}

When the external stress is less than the critical one, $\sigma<\sigma_c$ we
 use the positive parameter
\begin{equation}
\epsilon = \sigma_c - \sigma
\label{eq:2.4.5.1}\end{equation}
as a measure of the deviation from criticality. In this case the bundle is 
expected to relax to an equilibrium situation with a finite number of fibers 
intact. We are going to calculate how many iteration $t_f$ are needed to reach 
equilibrium.

Here we consider the uniform threshold distribution (\ref{eq:uniform}), and 
again we transform the nonlinear iteration (\ref{eq:2.4.1.12}) to a linear one 
by means of two transformations. Introducing
$\sigma= {\textstyle \frac{1}{4}}-\epsilon$ and 
\begin{equation}
n_t={\textstyle \frac{1}{2}}+\sqrt{\epsilon}\;/z_t 
\label{eq:2.4.5.2}\end{equation}
into (\ref{eq:2.4.1.12}), we have
\begin{equation}
2\sqrt{\epsilon} = \frac{z_{t+1}-z_t}{1-z_{t+1}z_t}.
\label{eq:2.4.5.3}\end{equation}
A second transformation,
\begin{equation}
z_t=\tanh w_t,
\label{eq:2.4.5.4}
\end{equation}
gives
\begin{equation}
2\sqrt{\epsilon} = \frac{\tanh w_{t+1}-\tanh w_t}{1-\tanh w_{t+1}\tanh w_t} \equiv \tanh (w_{t+1}-w_t).
\label{eq:2.4.5.5}\end{equation}
Hence we have the linear iteration $w_{t+1}-w_t= \tanh^{-1} (2\sqrt{\epsilon})$, which gives
\begin{equation}
w_t = w_0+t\;\tanh^{-1}(2\sqrt{\epsilon}).
\label{eq:2.4.5.6}\end{equation} 
As $\tanh^{-1}x=\frac{1}{2}\,\ln[(1+x)/(1-x)]$,  
via (\ref{eq:2.4.5.2}) and (\ref{eq:2.4.5.4}) the initial situation with no broken fibers, $n_0=1$, corresponds to $w_0=\tanh^{-1}(2\sqrt{\epsilon})$, so that (\ref{eq:2.4.5.6}) becomes
\begin{equation}
w_t=(1+t)\;\tanh^{-1} (2\sqrt{\epsilon}).
\label{eq:2.4.5.7}\end{equation}
For the original variable this corresponds to
\begin{equation}
n_t = \frac{1}{2}+\frac{\sqrt{\epsilon}}{\tanh[(1+t)\tanh^{-1}(2\sqrt{\epsilon})]}.
\label{eq:2.4.5.8}\end{equation}
 After an infinite number of iterations ($t\rightarrow \infty$ in (\ref{eq:2.4.5.8})) $n_t$  apparently approaches the fixed point 
\begin{equation}
n^* = {\textstyle \frac{1}{2}}+\sqrt{\epsilon},
\label{eq:2.4.5.9}\end{equation}
which is the fixed point (\ref{eq:2.4.1.10}) for the uniform distribution. However, the bundle contains merely a finite number of fibers, so equilibrium should be reached after a {\it finite} number of steps. Since an equilibrium value corresponds to a fixed point, we seek fixed points $N^*$ for finite $N$.   

\begin{figure}[h]
\begin{center}
\includegraphics[width=8cm]{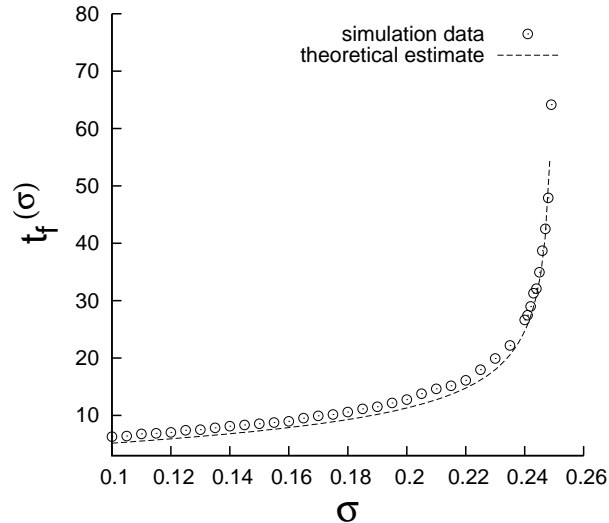}
\caption {\label{fig:pre-crit}
Pre-critical relaxation: Numerical data are for a bundle with $N=10^6$ fibers
having uniform threshold distribution and averages are taken over 
$10^5$ samples.}
\end{center}
\end{figure}

Taking into account that the variables $N_t$ are integers, one can get the final result (see \cite{ph07,hhp15}) 
\begin{equation}
t_f(\sigma) = -1 +\frac{\ln (N)}{2\tanh^{-1}(2\sqrt{\epsilon})}.
\label{eq:2.4.5.14}\end{equation} 
The simulation data in Figure (\ref{fig:pre-crit}) for the uniform threshold 
distribution are well approximated by this analytic formula.
Equation (\ref{eq:2.4.5.14}) shows that near the critical point the number of iterations has again a square root divergence, 
\begin{equation}
t_f = {\textstyle \frac{1}{4}}\;\ln (N)\,(\sigma_c-\sigma)^{-1/2},
\label{eq:2.4.5.15}\end{equation}
with a system-size-dependent amplitude.

\end{document}